%% file: paper.tex
\documentclass[aps,prl,twocolumn,showpacs,superscriptaddress,preprintnumbers,floatfix]{revtex4}  % for review and submission 
\usepackage{graphicx}  % needed for figures
\usepackage{dcolumn}   % needed for some tables
\usepackage{bm}        % for math
\usepackage{amssymb}   % for math
\usepackage{xspace}
\usepackage{multirow}
\usepackage{color}
\usepackage{units}
\newcommand{\pot}{\ensuremath{\textrm{POT}}\xspace}

\newcommand{\fullnopot}{\ensuremath{2.95\times10^{20}}\xspace}
\newcommand{\rIVpot}{\ensuremath{1.71\times10^{20}\,\pot}\xspace}
\newcommand{\rVIIpot}{\ensuremath{1.24\times10^{20}\,\pot}\xspace}
\newcommand{\baseline}{\unit[735]{km}}

\newcommand{\dmsq}[1]{\ensuremath{\Delta m^2_{ #1 }}}

\newcommand{\numu}{\ensuremath{\nu_{\mu}}\xspace}          % nu_mu
\newcommand{\numubar}{\ensuremath{\overline{\nu}_{\mu}}\xspace}   % nu_mu
\newcommand{\numucc}{\ensuremath{\nu_{\mu}}-CC\xspace} 
\newcommand{\numubarcc}{\ensuremath{\overline{\nu}_{\mu}}-CC\xspace}   % nu_mu

\newcommand{\nue}{\ensuremath{\nu_{e}}\xspace}                      % nu_e
\newcommand{\nutau}{\ensuremath{\nu_{\tau}}\xspace}              % nu_tau
\newcommand {\nubar}{\ensuremath{\bar{\nu}}}

\newcommand{\minos}{MINOS}
\newcommand{\pmns}{\ensuremath{U_{\mathrm{PMNS}}}}
\newcommand{\pmerr}[2]{\ensuremath{^{+#1}_{-#2}}}
\newcommand{\dm}{\ensuremath{\Delta m^{2}}}
\newcommand{\dmbar}{\ensuremath{ \Delta \overline{m}^{2} }\xspace}
\newcommand{\dmbarfull}{\ensuremath{\unit[\lbrack 2.62\pmerr{0.31}{0.28}\mbox{(stat.)} \pm 0.09 \mbox{(syst.)}\rbrack]{\times 10^{-3} eV^{2}}}}
\newcommand{\dmbarIV}{\ensuremath{\unit[\lbrack 3.36\pmerr{0.46}{0.40}\mbox{(stat.)} \pm 0.06 \mbox{(syst.)}\rbrack]{\times 10^{-3} eV^{2}}}}

\newcommand{\snbarIV}{\ensuremath{0.86\pm0.11 \mbox{(stat.)} \pm 0.01 \mbox{(syst.)} }}

\newcommand{\dmbarVII}{\ensuremath{\unit[\lbrack 2.26\pmerr{0.27}{0.29}\mbox{(stat.)} \pm 0.09 \mbox{(syst.)}\rbrack]{\times 10^{-3} eV^{2}}}}
\newcommand{\snbarlimitVII}{\ensuremath{0.79}\xspace}

\newcommand{\sn}{\ensuremath{\sin^{2}(2\theta)}}
\newcommand{\snbar}{\ensuremath{\sin^{2}(2\overline{\theta})}\xspace}
\newcommand{\snbarfull}{\ensuremath{0.95\pmerr{0.10}{0.11}\mbox{(stat.)} \pm 0.01 \mbox{(syst.)}}\xspace}
\newcommand{\snbarlimit}{\ensuremath{0.75}\xspace}

\newcommand{\noOscExp}{\ensuremath{273}\xspace}

\newcommand{\dmBestFit}{\ensuremath{2.32\times 10^{-3}eV^{2}}}

\newcommand{\snLimit}{\ensuremath{0.90}\xspace}

\newcommand{\dmfull}{\ensuremath{\unit[ 2.32\pmerr{0.12}{0.08}\mbox{(stat.+syst.)}]{\times 10^{-3} eV^{2}}}}

\newcommand{\muid}{\ensuremath{\mu\mathrm{ID}}\xspace}

\begin{document}

%\leftline{Version 6.0 as of \today} 
%\leftline{To be submitted to PRL}

\title{An improved measurement of muon antineutrino disappearance in MINOS}
%\title{A precision measurement of muon antineutrino oscillations with MINOS}

\input{authors.tex}

\date{Jan 19, 2012}
\preprint{FERMILAB-PUB-XX-YYY-E}
\preprint{BNL-XXXXX-YYYY-EE}
\preprint{arXiv:hep-ex/XXXX.XXXX}

\begin{abstract}
%We report measurements of muon antineutrino disappearance over a distance of \baseline{} using the Main Injector neutrino beam at Fermilab and the \minos{} detectors. The beam was configured to focus $\pi^{-}$ to provide a \numubar{} enhanced beam in the few GeV energy range where ``atmospheric'' oscillations occur.  With a total of exposure of \fullpot{}, of which \rVIIpot{} has not been previously analyzed, we find $\dmbar=\dmbarfull$ and $\snbar{}>\snbarlimit{}$ (90\%CL). These values supersede our previous measurements of \dmbar{} and \snbar{} and are in agreement with \dm{} and \sn{} measured in accelerator and atmospheric oscillation experiments.

%We report results from improved measurements of \numubar{} disappearance over a distance of \baseline{} using the \minos{} detectors and the Femilab NuMI beam in a \numubar{}-enhanced configuration.  From a total exposure of \fullpot{}, of which \rVIIpot{} have not been previously analyzed, we make the most precise measurement of  $\dmbar=\dmbarfull$ and further constrain the \numubar{} mixing angle $\snbar{}>\snbarlimit{}$ (90\%CL). These values are in agreement with \dm{} and \sn{} measured for \numu{} in accelerator and atmospheric experiments.

We report an improved measurement of \numubar{} disappearance over a distance of \baseline{} using the \minos{} detectors and the Fermilab Main Injector neutrino beam in a \numubar{}-enhanced configuration.  From a total exposure of \fullnopot{} protons on target, of which 42\% have not been previously analyzed, we make the most precise measurement of  $\dmbar=\dmbarfull$ and constrain the \numubar{} mixing angle $\snbar{}>\snbarlimit{}$ (90\%CL). These values are in agreement with \dm{} and \sn{} measured for \numu{}, removing the tension reported in~\cite{ref:minosRHC2011}.
\end{abstract}

\maketitle

%\section*{Introduction}

Observations of neutrinos and antineutrinos created in the Sun, the Earth's atmosphere, reactors and accelerators provide strong evidence~\cite{ref:minos2006, ref:minos2008, ref:minosCC2010, ref:sk, ref:soudan2, ref:macro, ref:k2k, ref:sno, ref:kamland} that neutrinos undergo transitions between their flavor eigenstates ($\nue$, $\numu$, $\nutau$) as they propagate. These transitions can occur due to quantum mechanical mixing between the neutrino flavor and mass ($\nu_1$, $\nu_2$, $\nu_3$) eigenstates.  The mixing may be parametrized with a unitary matrix \pmns{}~\cite{ref:PMNS} which is typically expressed in terms of three mixing angles $\theta_{12}$, $\theta_{23}$, $\theta_{13}$ and a charge-parity (CP) violating phase $\delta$.  This interpretation, referred to as ``neutrino oscillations,'' requires that neutrinos have mass and motivates extensions to the Standard Model (SM) of particle physics. Extensions which explain the origin of neutrino masses, for example the addition of right handed sterile neutrinos $\nu_{R}$~\cite{Mohapatra:1979ia},  may also explain the baryon asymmetry~\cite{Fukugita:1986hr} of the universe.

%dark matter puzzle and the
%~\cite{unknown}.

%For \nue{} and \nuebar{} traversing matter, the charged current scattering on atomic electrons introduces an apparent difference between \nue{} and \nuebar{} oscillations~\cite{ref:needed}. This ``matter effect'' may be exploited to measure $\delta$ and determine if $m_{3}>m_{2}, m_{1}$ or $m_{3}<m_{2}, m_{1}$~\cite{ref:needed}. Within the Standard Model, a analogous matter effect is prohibited for muon and tau neutrinos; the probability $P(\numu\rightarrow\numu)$ that a \numu{} produced via $\pi^{+}\rightarrow \mu^{+}\numu$ is detected after a distance $L$ as a \numu{} (rather than a \nue{} or \nutau{}) must be equal to the corresponding probability $P(\numubar\rightarrow\numubar)$ for antineutrinos.  Observation of $P(\numu\rightarrow\numu) \neq P(\numubar\rightarrow\numubar)$ would therefore be evidence for physics beyond the Standard Model.

%Even when the SM is extended to include $\nu_{R}$, 

The CPT symmetry of the SM requires that \numu{} and \numubar{} have the same masses and mixing parameters.  In vacuum, the probability $P(\numu\rightarrow\numu)$ that a \numu{} is detected after a distance $L$ as a \numu{} (rather than a \nue{} or \nutau{}) must be equal to the corresponding probability $P(\numubar\rightarrow\numubar)$ for antineutrinos. For a \numu{} with energy $E$ the probability may be written as
\begin{equation}
P(\numu\rightarrow\numu) = 1 - \sn\sin^{2}\left(\frac{1.267\dm\mathrm{[eV^{2}]} L\mathrm{[km]}}{E \mathrm{[GeV]}}\right)
\label{eqn:OscProb}
\end{equation}
where \dm{} and \sn{} are effective parameters that are functions of the angles parameterizing \pmns{} and the differences in the squared masses $\dmsq{ij}=m^{2}_i - m^{2}_j$ of the $\nu_1$, $\nu_2$ and $\nu_3$ states. Experiments have demonstrated $|\dmsq{31}|\gg |\dmsq{21}|$~\cite{ref:PDG}. In the limiting case that $\theta_{13}\approx 0$ we have $\theta\approx\theta_{23}$ and $|\dm|\approx\sin^{2}(2\theta_{12})|\dmsq{31}|+\cos^{2}(2\theta_{12})|\dmsq{32}|$~\cite{Parke:2008zz,ref:kamland2}.  Muon antineutrino oscillations are described by an equation which has the same form as Eq.~\ref{eqn:OscProb} with parameters \dmbar and \snbar.  The extended SM predicts $\dmbar=\dm$ and $\snbar=\sn$ for vacuum oscillations~\cite{ref:mattereffect,Wolfenstein:1977ue}. Observation of $P(\numu\rightarrow\numu) \neq P(\numubar\rightarrow\numubar)$ would therefore be evidence for physics beyond the SM, such as neutrino interactions in the earth's crust that do not conserve lepton flavor.

%Comparisons of $P(\numu\rightarrow\numu)$ and $P(\numubar\rightarrow\numubar)$ therefore provide a sensitive test of neutrino oscillations in the SM and act as a constraint on new physics.

In this Letter we describe a measurement of $P(\numubar\rightarrow\numubar)$ conducted over a baseline $L=\baseline$ using a \numubar{}-enhanced beam with a peak energy of \unit[3]{GeV}. The beam was produced by directing \unit[120]{GeV/c} protons from the Fermilab Main Injector onto a graphite target to produce $\pi/K$ mesons that decay to produce neutrinos. Two magnetic horns focus the mesons, allowing us to control the energy spectrum and $\nu$/$\nubar$ content of the beam.

The neutrino beam is pointed towards two detectors, referred to as Near and Far. The \unit[980]{ton} Near Detector (ND) measures the \numu{} and \numubar{} content of the beam as a function of energy at a distance of \unit[1.04]{km} from the $\pi/K$ production target. The \unit[5.4]{kton} Far Detector (FD) is located in the Soudan Underground Laboratory, $\unit[734]{km}$ from the ND, and remeasures the beam composition. The neutrino detectors are steel-scintillator, tracking-sampling calorimeters optimized to identify and measure the energy of muon neutrinos and antineutrinos and reject backgrounds from neutral current and \nue{} interactions~\cite{ref:minosnim}. The detectors are magnetized with an average field of \unit[1.3]{T} to distinguish \numu from \numubar based on the charge of the $\mu$ produced in weak interactions.

We previously reported \numu{} oscillations with an energy dependence consistent with Eq.~\ref{eqn:OscProb} and $\dm=\dmfull$, $\sn>\snLimit$~(90\% CL)~\cite{ref:minosCC2010}. The measurements utilized $7.25\times10^{20}$ protons on target (\pot{}) of data collected between 2005-2009 with a \numu{}-enhanced beam~\cite{ref:beamcomp}. Measurements made by Super-Kamiokande~\cite{ref:sk} and T2K~\cite{ref:t2k} are in good agreement with our values.

%These values, those made by Super-Kamiokande using atmospheric neutrinos~\cite{ref:sk}, and measurements by T2K using an accelerator beam~\cite{ref:t2k} are in good agreement.

In 2009-2010 we collected \rIVpot{} in a \numubar{}-enhanced beam~\cite{ref:beamcomp} created by reversing the polarity of the horns. The magnetic fields in the FD and ND were also reversed to focus the $\mu^{+}$ created in \numubar{} interactions. These antineutrino data also exhibited oscillations in agreement with Eq.~\ref{eqn:OscProb}, but with parameters $\dmbar=\dmbarIV$ and $\snbar=\snbarIV$~\cite{ref:minosRHC2011}. We found no systematic effects which could explain the difference between the \numu and \numubar parameters. Assuming identical true values for $\left(\dmbar,\snbar\right)$ and $\left(\dm,\sn\right)$, we calculated that such a difference would occur by random chance about 2\% of the time.  To clarify the situation we collected an additional \rVIIpot{} with the \numubar{}-enhanced beam during 2010-2011. We have also updated the analysis to improve the sensitivity to $\left(\dmbar,\snbar\right)$, reduce uncertainties due to Monte Carlo (MC) modeling in the ND, and increase the similarity to the \numu{} oscillation analysis.

%A total of $2.98\times10^{20}$ protons were delivered to the neutrino target during the data-taking periods. We impose a series of data and beam quality~\cite{ref:minos2008,ref:beam,ref:beam2} requirements which reduce the analyzable exposure to $\unit[2.95\times 10^{20}]{POT}$ (99.0\% livetime) at the FD, and $\unit[2.73\times 10^{20}]{POT}$ (91.8\% livetime) at the ND. The uncertainty in the livetime is negligible, and the high and largely overlapping livetime in both detectors assures that our results are not sensitive to beam effects that would cause the number of events per \pot{} to vary.

%\section*{Event Selection and Reconstruction}

\begin{figure}
\begin{center}
\includegraphics[width=\columnwidth]{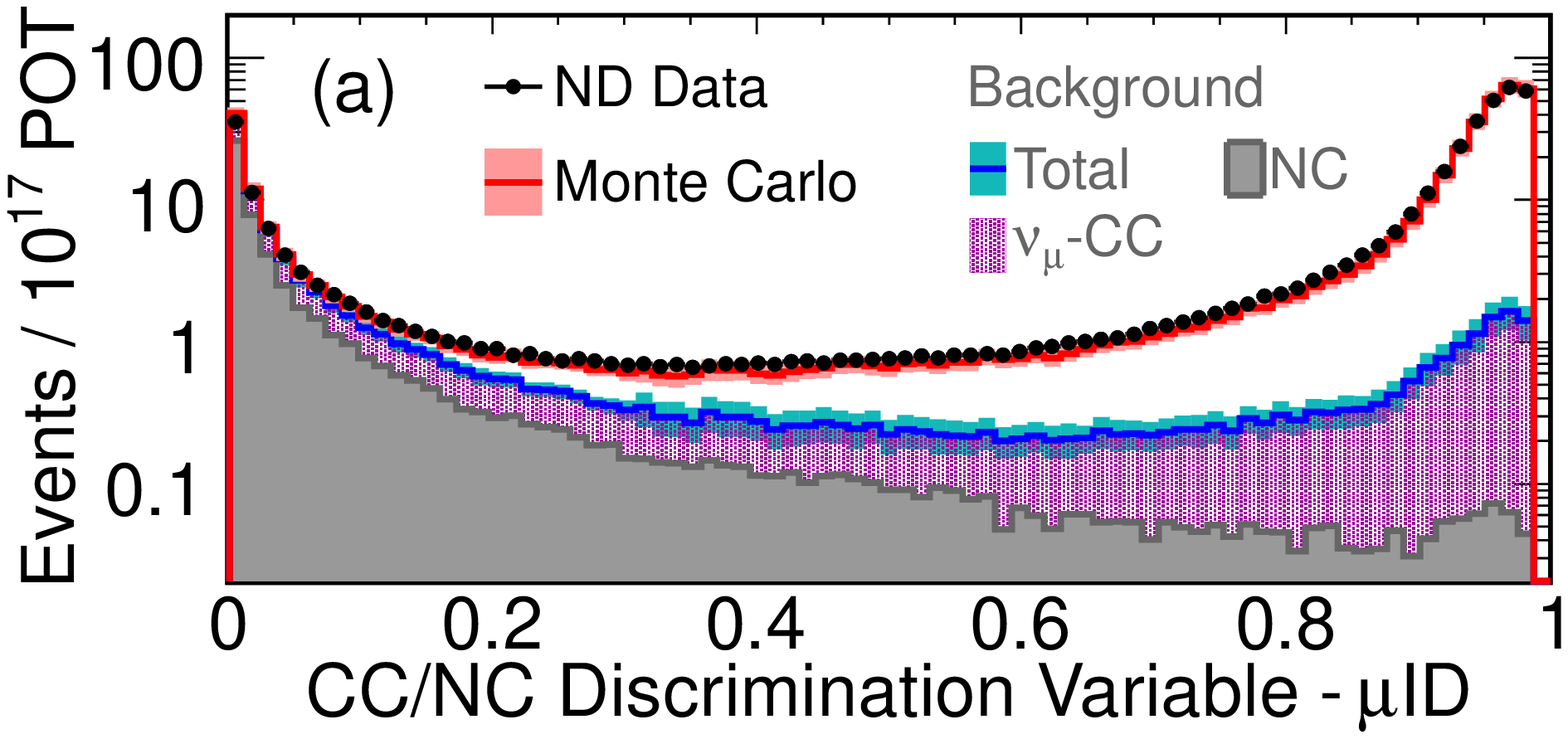}
%\includegraphics[width=\columnwidth]{new}
%\rule{\columnwidth}{1pt}
%\includegraphics[width=\columnwidth]{hqp_sigqpN_1}
\includegraphics[width=\columnwidth]{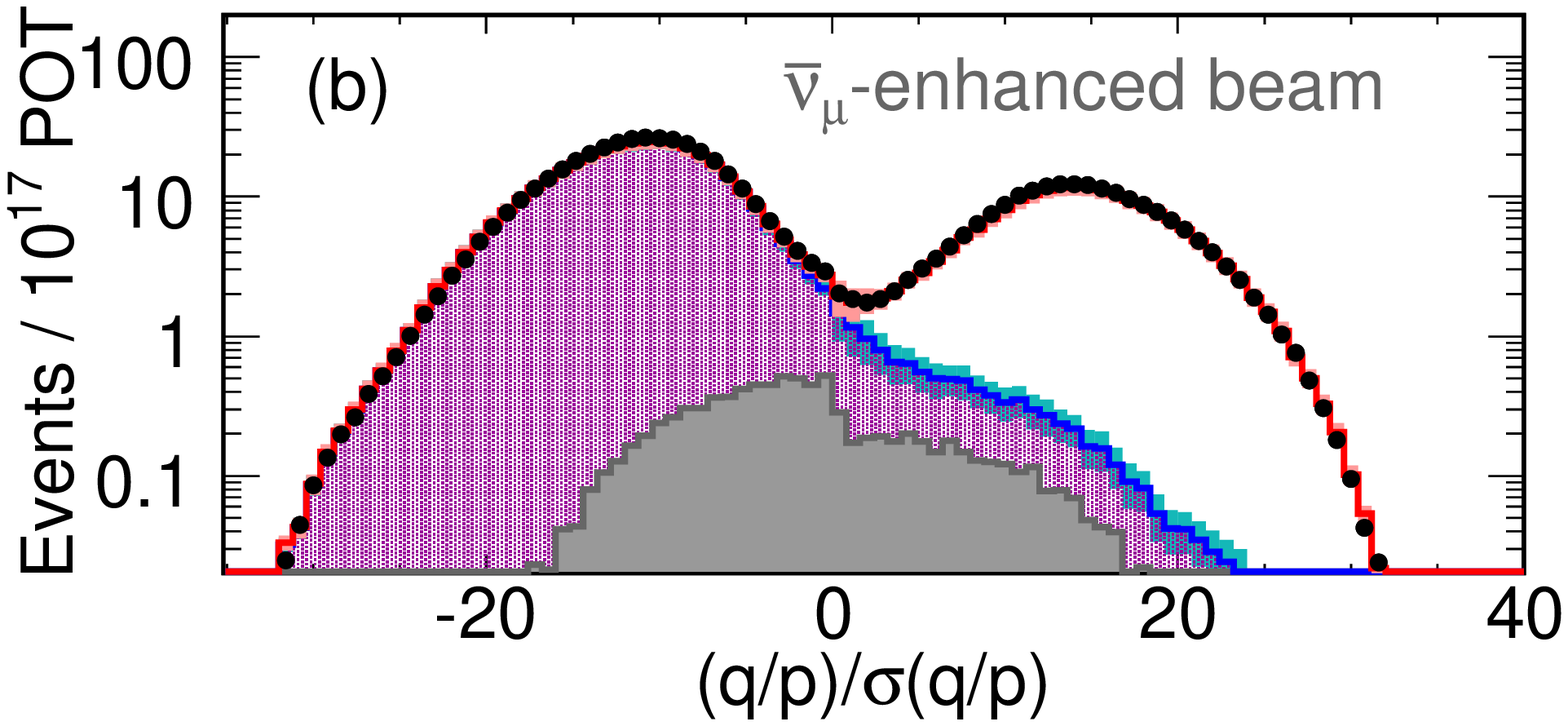}

\end{center}
\caption{\label{fig:pid} Event selection variables from the ND: (a) The variable, \muid{}, used to select \numu{}/\numubarcc{} events and reject NC events. (b) The reconstructed track charge/momentum $q/p$ divided by the uncertainty $\sigma(q/p)$ reported by the track reconstruction algorithm. In both figures, we have applied all selection criteria except the one on the quantity being shown. Shaded bands show systematic uncertainties. The background histograms are stacked on top of each other.}
\end{figure}

\begin{figure}
\begin{center}
\includegraphics[width=\columnwidth]{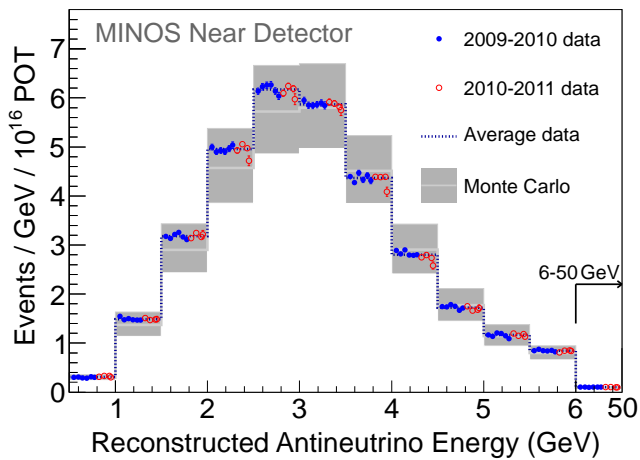}
\end{center}
\caption{The \numubarcc{} energy spectrum measured by the ND. The selection procedure described in the text has been applied. Each point represents one month of data. The shaded Monte Carlo band shows the combined effect of systematic uncertainties due to cross-sections, flux, energy scale and other sources.}
\label{fig:ndspec}
\end{figure}

 We isolate a sample of \numu~and \numubar~charged-current (CC) $\numu{} N \rightarrow \mu X$ events by searching for interaction vertices inside our detectors with a muon track and possible hadronic activity from the recoil system $X$. We reject hadron tracks reconstructed in neutral-current (NC) events by combining four topological variables describing track properties into a single discriminant variable, \muid{}, using a $k$-nearest-neighbor (kNN) technique~\cite{ref:RustemThesis}. The kNN algorithm calculates the distance in the four-variable space between each measured event and an ensemble of simulated events; the output is the fraction of signal in the $k=80$ closest MC events. This discriminant was used in our previous analyses~\cite{ref:minosCC2010,ref:minosRHC2011} and, as shown in Fig.~\ref{fig:pid}(a), is well modeled by our MC. We maximize the statistical sensitivity to \dmbar by requiring $\muid>0.3$, which results in a MC estimated efficiency/purity of 90.7\%/99.0\% at the ND and 91.6\%/99.0\% at the FD. We then discriminate \numu{} from \numubar{} on an event-by-event basis by analyzing the track curvature in the detector's magnetic field.  Figure~\ref{fig:pid}(b) shows the track charge/momentum ($q/p$) divided by its uncertainty ($\sigma(q/p)$), as determined by our track reconstruction algorithm. We select $\numubar{}$ and reject \numu{} by requiring $\frac{q/p}{\sigma(q/p)}>0$ with a MC estimated efficiency/purity of 98.4\%/94.7\% in the ND and 98.8\%/95.1\% in the FD. The \numu background accepted by the selection is predominantly due to high energy muons with small curvature. 

We reconstruct the neutrino energy by summing muon and hadronic shower energies.  The muon energy is measured using track range and curvature.  We reconstruct the hadronic shower energy using three variables: the sum of the reconstructed energy deposited by showers that start within \unit[1]{m} of the track vertex; the sum of the energy in the two largest showers reconstructed in the event; and the length of the longest shower. We use these three variables in a second kNN algorithm and estimate the shower energy as the mean true hadronic energy of the $k=400$ closest MC events. This technique improves the hadronic energy resolution when compared to a method which uses only the energy deposited by the largest shower, increases the statistical sensitivity to \dmbar~by 10\%, and was previously used to analyze \numu{} disappearance~\cite{ref:ChrisThesis,ref:minosCC2010}.

Data from the ND are used to predict the neutrino energy distribution at the FD. Though both detectors have the same segmentation and very similar average magnetic fields, for economic reasons the ND is smaller and asymmetric about the magnetic field coil and is more coarsely instrumented with scintillator in the downstream ``muon spectrometer'' region~\cite{ref:minosnim}. In addition, the ND coil occupies a larger fractional area than the FD coil and more muons enter it. In the ND data, we observe a reconstruction failure rate of 6.1\%, mostly associated with tracks entering the coil region, but the MC predicts 4.2\%. Previously, we dealt with this issue by assigning a systematic error. Now, we remove ND events with a track that ends less than \unit[60]{cm} from the coil. We also remove events with a track that ends on the side of the coil opposite the beam centroid. The new event selection decreases the efficiency to 53\% in the ND, but reduces the data/MC failure rates to 1.4\%/0.9\%. The selected sample contains the same classes of neutrino scattering processes as are present at the FD, and our results are not significantly more vulnerable to cross-section uncertainties. We applied the new selection and shower energy reconstruction to the 2009-2010 data and found that the best fit parameters shifted by only $\delta(\dmbar)=\unit[+1.0\times 10^{-4}]{eV^2}$ and $\delta(\snbar)=-3.6\times10^{-2}$.

A total of $2.98\times10^{20}$ protons were delivered to the graphite target during the data-taking periods. We impose a series of data and beam quality~\cite{ref:minos2008,ref:beam,ref:beam2} requirements which reduce the analyzable exposure to $\unit[2.95\times 10^{20}]{POT}$ (99.0\% livetime) at the FD, and $\unit[2.73\times 10^{20}]{POT}$ (91.8\% livetime) at the ND. The uncertainty in the livetime is negligible, and the high and largely overlapping livetime in both detectors assures that our results are not sensitive to beam effects that would cause the number of events per \pot{} to vary.

Under normal conditions, the ND measures about 2400 \numubarcc{} events per day in the oscillation energy region ($E_\nu<\unit[6]{GeV}$). These data are essential for monitoring the neutrino beam and the quality of the experiment. Figure~\ref{fig:ndspec} shows the reconstructed \numubar{} energy distribution measured in each month during the data-taking periods. With the exception of Feb.~2011 (the last point in each bin in Fig.~\ref{fig:ndspec}), all months are in statistical agreement, and we expect a constant counting rate per \pot{} at the FD. Part of the February dataset was taken after the neutrino target's cooling system had failed, leaking water into the target canister. This resulted in a decrease in the neutrino flux of 4\% from 0-\unit[6]{GeV} when integrated over the entire month. The decrease is adequately modeled by our beam simulation, and we account for it at the FD using the ND data.

%and using the ND data it is accounted for at the FD with negligible uncertainty.

%The data from inline monitors which measure

Measurements of the beam position and width at the target, the position of the remnant proton beam at the end of the decay pipe, and the integrated muon flux from meson decays in the pipe, all further indicate that the expected number of \numubarcc{} per POT is constant at the FD. Using the proton beam and monitors we measured the target and horn misalignments as well as the residual magnetic field in the neck of the focusing horns~\cite{ref:horns}. We conclude that these effects introduce $\ll 1\%$ uncertainty in the rate at the FD.

\begin{figure}
\begin{center}
\includegraphics[width=\columnwidth]{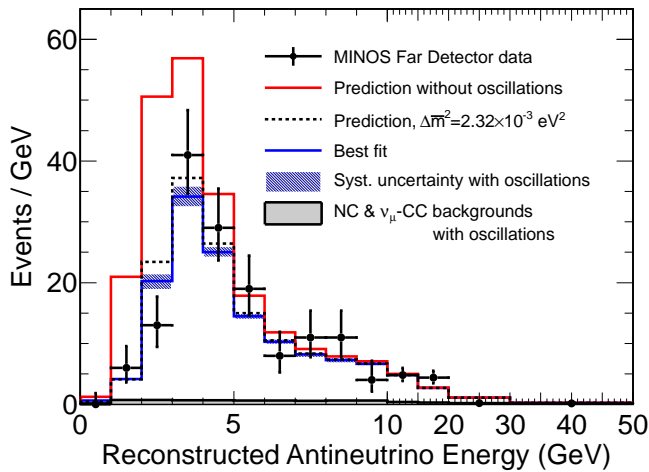}
\end{center}
\caption{\label{fig:fdspec} The reconstructed energy of FD \numubarcc{} events. These data are fit to produce the 2009-2011 contour shown in Fig.~\ref{fig:cont}. The band displays the effect of systematic uncertainties. }

\end{figure}

Events at the FD are read out in a $\unit[100]{\mu s}$ window surrounding the $\unit[10]{\mu s}$ long beam spill~\cite{ref:spilltrigger}. We select events coincident with the beam spill that have a vertex in the fiducial volume and a track identified as a muon by \muid{}. We remove cosmic-ray background by requiring that the cosine of the angle between the track and beam direction be $>0.6$.  A total of 521 events satisfy our criteria, 273 in the 2009-2010 dataset and 248 in 2010-2011. Using the muon charge, we identify 328 as \numu{} and 193 as \numubar{}. We apply the selection to $37.7\times 10^7$  readout windows taken in anticoincidence with the beam during 2005-2011 and estimate a cosmic-ray background of 0.8 events.  We monitor detector and reconstruction performance by measuring the rate of cosmic-ray muons traversing the detector before and after the beam spill. The rates in each data-taking period, $\unit[0.38\pm0.03]{Hz}$ in 2009-2010 and $\unit[0.41\pm0.04]{Hz}$ in 2010-2011, are consistent.

%These cosmic-rays occur at a rate of $\unit[0.40\pm0.03]{Hz}$, consistent between the data collected in 2009-2010 and the new 2010-2011 data.

We predict the \numubar{} energy spectrum at the FD by first correcting the ND spectrum for inefficiency and backgrounds. We then transfer that spectrum to the FD using a two dimensional ``beam matrix''~\cite{ref:minos2008}. We predict \noOscExp events if \numubar{} do not oscillate, including 3.5 NC and 10.8 \numucc{}~\footnote{\numu are assumed to oscillate with $\dm=\dmBestFit$ and $\sn=1$.}. The energy spectra are shown in Fig.~\ref{fig:fdspec}. Oscillations are incorporated into the prediction according to Eq.~\ref{eqn:OscProb}. Maximizing the binned log-likelihood yields
\begin{eqnarray*}
\dmbar&=&\dmbarfull \\
\snbar&=&\snbarfull \\
\snbar&>&\snbarlimit \quad \mbox{(90\% C.L.)}
\end{eqnarray*}
with $p=35.3\%$ at best fit~\footnote{The ``p-value'', $p$, is the probability, assuming our fit hypothesis is true, of obtaining data more incompatible with our hypothesis than the data we actually observe. See the discussion in section 33.2.2 of~\cite{ref:PDG}.}. When analyzing the 2010-11 alone we obtain  $\dmbar=\dmbarVII$, $\snbar>\snbarlimitVII$ (90\% C.L.) with $p=14.5\%$. 

The systematic errors stated above were evaluated by computing the standard deviation in the best fit parameters as the fits were repeated using MC samples that were shifted in accordance with uncertainties in neutrino cross-sections; the beam matrix and relative FD normalization; NC and \numucc backgrounds; the relative FD to ND energy calibration; the absolute muon energy scale; and the absolute hadronic energy scale, including final state hadronic interactions. The input uncertainties are as in our previous analyses~\cite{ref:minosRHC2011,ref:minosCC2010} and were derived from in-situ data, bench tests of detector and beam components, a test beam experiment, and published neutrino and hadron cross-sections~\cite{ref:minos2008,ref:neugen}. We increased the uncertainty in the axial-vector mass from 15\% to 30\% to account for additional uncertainties in \numubar quasi-elastic scattering~\cite{ref:MesonExchangeCurrents}. 

%The systematic errors stated above were evaluated by computing the standard deviation in the best fit parameters as the fits were repeated using MC samples that were shifted in accordance with the following systematic uncertainties:\\
w
%\begin {tabular} {l c r }
%\hline
%Source   of & $\rm \delta (\delmsq{})$  &  $\rm \delta (sin^2(2\theta))$\\
%systematic uncertainty                 & $\rm (10^{-3}\,eV^2)$       &  \\
%\hline
%hadronic energy scale & 0.057 &  0.006\\
%muon energy scale & 0.056 &      0.002\\                       
%relative FD/ND normalization & 0.024 &  0.002\\
%NC contamination & 0.012 &      0.007\\
%neutrino cross-sections  & 0.019 &      0.006 \\
%all others               & 0.007 &      0.001   \\
%\hline
%\end {tabular}
%\\
%The input uncertainties are as in our previous analyses~\cite{ref:minosRHC2011,ref:minosCC2010,ref:minos2008,ref:neugen}. We increased the uncertainty in the axial-vector mass from 15\% to 30\% to account for additional uncertainties in \numubar quasi-elastic scattering~\cite{ref:MesonExchangeCurrents}. 

\begin{figure}
\begin{center}
\includegraphics[width=1.0\columnwidth]{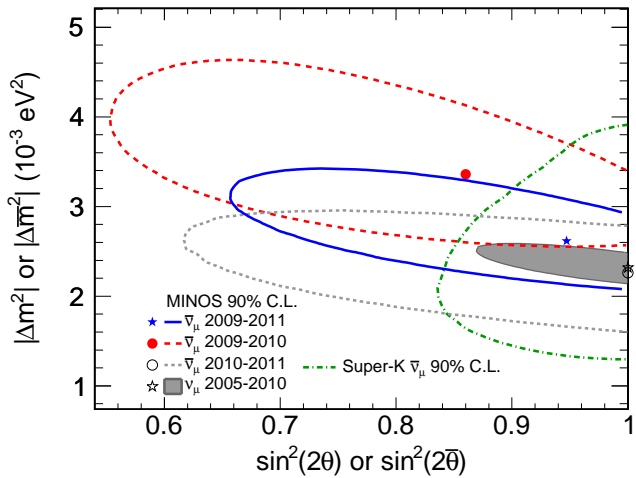}
\end{center}
\caption{Confidence regions calculated by fitting the FD data with an energy spectrum predicted from the ND but modified to incorporate oscillations via  Eq.~\ref{eqn:OscProb}. The regions include statistical and systematic uncertainties. The $\numubar$ 2009-2011 contour is derived from the energy spectrum shown in Fig.~\ref{fig:fdspec}.}
\label{fig:cont}
\end{figure}

% old version of the FC discussion
%The confidence regions shown in Fig.~\ref{fig:cont} incorporate both statistical and systematic uncertainties. Following the Feldman-Cousins prescription~\cite{ref:FC}, we divided the $(\dmbar,\snbar)$ space in a fine grid~\footnote{The grid granularity was $(\unit[5\times10^{-5}]{eV^{2}},0.02$)} and simulated 100,000 experiments at each point $(\dmbar_i,\snbar_i)$. For each experiment, we calculated the likelihood assuming (a) $(\dmbar_i,\snbar_i)$ and (b) the best fit parameters found when analyzing the data. The confidence level at the grid point is the fraction of experiments that give a larger likelihood for  $(\dmbar_i,\snbar_i)$ than for the best fit. Systematic uncertainties are incorporated into the confidence regions by varying the systematic error sources discussed above when generating the experiments.

% through the same procedure: each fake experiment is simulated using a different amount of mismodeling for each source of uncertainty. The amount of mismodeling is chosen randomly, in accordance with the estimated $1\sigma$ uncertainties.

%This shows the previous \numubar measurement made by MINOS to be consistent with a statistical fluctuation.

%and the fit quality is acceptable: a fake data study shows that 14.5\% of experiments are expected to have a worse fit than the data. 

% Do I go back to the tension here?  Can't make it fit. Maybe the wrong choice.
 
The confidence regions shown in Fig.~\ref{fig:cont} were calculated according to the unified procedure of~\cite{ref:FC} and incorporate both statistical and systematic uncertainties. We assess the consistency of the \numu and \numubar measurements by doing a joint $(\dmbar=\dm,\snbar=\sn)$ fit to the data to establish the null hypothesis. We then perform four-parameter $(\dmbar,\dm,\snbar,\sn)$ fits on an ensemble of \numu{} and \numubar{} MC experiments generated at the joint best fit. The joint fit had a larger likelihood than $p=42\%$ of the four parameter fits, indicating consistency between \numu{} and \numubar{}. 

%This indicates a high level of consistency between \numu and \numubar disappearance for $L/E\approx\unit[200]{km/GeV}$.

In conclusion, we have used a \numubar{}-enhanced Fermilab accelerator beam and detectors that discriminate \numu from \numubar to make the most precise measurement of \dmbar{}. Our results remove the tension reported in~\cite{ref:minosRHC2011} and establish consistency between \numu{} and \numubar{} oscillations at $L/E\approx\unit[200]{km/GeV}$.

This work was supported by the U.S. DOE; the United Kingdom STFC; the U.S. NSF; the State and University of Minnesota; the University of Athens, Greece; Brazil's FAPESP, CNPq and CAPES; and the Jeffress Memorial Trust.  We are grateful to the Minnesota Department of Natural Resources and the personnel of the Soudan Laboratory and Fermilab.

%This work was supported by the U.S. DOE; the United Kingdom STFC; the U.S. NSF; the State and University of Minnesota; the University of Athens, Greece; and Brazil's FAPESP and CNPq.  We are grateful to the Minnesota Department of Natural Resource and the personnel of the Soudan Laboratory and Fermilab.

%*********************************************************************72
%*********************************************************************72
%*********************************************************************72

%\begin{table}
%\begin{tabular}{l l c}
%\hline 
%\multirow{2}{*}{Uncertainty source} & & Uncertainty on\\
% & &  background events \\
%\hline 
%Far/Near ratio: &&4.5\%\\
%\ \ (a) Energy Scale &2.8\%&\\
%\ \ (b) Relative Event Rate  & 2.4\%&\\
%\ \ (c) Hadronic Model &2.5\%&\\
%\ \ (d) All Other Combined & 0.7\%&\\
%\hline
%Near Detector Decomposition & &2.8\% \\
%\nutau{} background & &1.7\%\\
%\hline

%Total Systematic Uncertainty && 5.6\% \\
%Expected Statistical Uncertainty && 14.3\% \\
%\hline
%\end{tabular}
%\caption{Systematic uncertainty in the total number of background events in the far detector.}
%\label{tab:systematics}
%\end{table}

\bibliography{paper}

\end{document}

%% file: authors.tex
\newcommand{\Berkeley}{Lawrence Berkeley National Laboratory, Berkeley, California, 94720 USA}
\newcommand{\Cambridge}{Cavendish Laboratory, University of Cambridge, Madingley Road, Cambridge CB3 0HE, United Kingdom}
\newcommand{\FNAL}{Fermi National Accelerator Laboratory, Batavia, Illinois 60510, USA}
\newcommand{\RAL}{Rutherford Appleton Laboratory, Science and Technologies Facilities Council, OX11 0QX, United Kingdom}
\newcommand{\UCL}{Department of Physics and Astronomy, University College London, Gower Street, London WC1E 6BT, United Kingdom}
\newcommand{\Caltech}{Lauritsen Laboratory, California Institute of Technology, Pasadena, California 91125, USA}
\newcommand{\Alabama}{Department of Physics and Astronomy, University of Alabama, Tuscaloosa, Alabama 35487, USA}
\newcommand{\ANL}{Argonne National Laboratory, Argonne, Illinois 60439, USA}
\newcommand{\Athens}{Department of Physics, University of Athens, GR-15771 Athens, Greece}
\newcommand{\NTUAthens}{Department of Physics, National Tech. University of Athens, GR-15780 Athens, Greece}
\newcommand{\Benedictine}{Physics Department, Benedictine University, Lisle, Illinois 60532, USA}
\newcommand{\BNL}{Brookhaven National Laboratory, Upton, New York 11973, USA}
\newcommand{\CdF}{APC -- Universit\'{e} Paris 7 Denis Diderot, 10, rue Alice Domon et L\'{e}onie Duquet, F-75205 Paris Cedex 13, France}
\newcommand{\Cleveland}{Cleveland Clinic, Cleveland, Ohio 44195, USA}
\newcommand{\Delhi}{Department of Physics \& Astrophysics, University of Delhi, Delhi 110007, India}
\newcommand{\GEHealth}{GE Healthcare, Florence South Carolina 29501, USA}
\newcommand{\Harvard}{Department of Physics, Harvard University, Cambridge, Massachusetts 02138, USA}
\newcommand{\HolyCross}{Holy Cross College, Notre Dame, Indiana 46556, USA}
\newcommand{\IIT}{Department of Physics, Illinois Institute of Technology, Chicago, Illinois 60616, USA}
\newcommand{\Iowa}{Department of Physics and Astronomy, Iowa State University, Ames, Iowa 50011 USA}
\newcommand{\Indiana}{Indiana University, Bloomington, Indiana 47405, USA}
\newcommand{\ITEP}{High Energy Experimental Physics Department, ITEP, B. Cheremushkinskaya, 25, 117218 Moscow, Russia}
\newcommand{\JMU}{Physics Department, James Madison University, Harrisonburg, Virginia 22807, USA}
\newcommand{\LASL}{Nuclear Nonproliferation Division, Threat Reduction Directorate, Los Alamos National Laboratory, Los Alamos, New Mexico 87545, USA}
\newcommand{\Lebedev}{Nuclear Physics Department, Lebedev Physical Institute, Leninsky Prospect 53, 119991 Moscow, Russia}
\newcommand{\LLL}{Lawrence Livermore National Laboratory, Livermore, California 94550, USA}
\newcommand{\LosAlamos}{Los Alamos National Laboratory, Los Alamos, New Mexico 87545, USA}
\newcommand{\MIT}{Lincoln Laboratory, Massachusetts Institute of Technology, Lexington, Massachusetts 02420, USA}
\newcommand{\Minnesota}{University of Minnesota, Minneapolis, Minnesota 55455, USA}
\newcommand{\Crookston}{Math, Science and Technology Department, University of Minnesota -- Crookston, Crookston, Minnesota 56716, USA}
\newcommand{\Duluth}{Department of Physics, University of Minnesota -- Duluth, Duluth, Minnesota 55812, USA}
\newcommand{\Ohio}{Center for Cosmology and Astro Particle Physics, Ohio State University, Columbus, Ohio 43210 USA}
\newcommand{\Otterbein}{Otterbein College, Westerville, Ohio 43081, USA}
\newcommand{\Oxford}{Subdepartment of Particle Physics, University of Oxford, Oxford OX1 3RH, United Kingdom}
\newcommand{\PennState}{Department of Physics, Pennsylvania State University, State College, Pennsylvania 16802, USA}
\newcommand{\PennU}{Department of Physics and Astronomy, University of Pennsylvania, Philadelphia, Pennsylvania 19104, USA}
\newcommand{\Pittsburgh}{Department of Physics and Astronomy, University of Pittsburgh, Pittsburgh, Pennsylvania 15260, USA}
\newcommand{\IHEP}{Institute for High Energy Physics, Protvino, Moscow Region RU-140284, Russia}
\newcommand{\Rochester}{Department of Physics and Astronomy, University of Rochester, New York 14627 USA}
\newcommand{\RoyalH}{Physics Department, Royal Holloway, University of London, Egham, Surrey, TW20 0EX, United Kingdom}
\newcommand{\Carolina}{Department of Physics and Astronomy, University of South Carolina, Columbia, South Carolina 29208, USA}
\newcommand{\SLAC}{Stanford Linear Accelerator Center, Stanford, California 94309, USA}
\newcommand{\Stanford}{Department of Physics, Stanford University, Stanford, California 94305, USA}
\newcommand{\StJohnFisher}{Physics Department, St. John Fisher College, Rochester, New York 14618 USA}
\newcommand{\Sussex}{Department of Physics and Astronomy, University of Sussex, Falmer, Brighton BN1 9QH, United Kingdom}
\newcommand{\TexasAM}{Physics Department, Texas A\&M University, College Station, Texas 77843, USA}
\newcommand{\Texas}{Department of Physics, University of Texas at Austin, 1 University Station C1600, Austin, Texas 78712, USA}
\newcommand{\TechX}{Tech-X Corporation, Boulder, Colorado 80303, USA}
\newcommand{\Tufts}{Physics Department, Tufts University, Medford, Massachusetts 02155, USA}
\newcommand{\UNICAMP}{Universidade Estadual de Campinas, IFGW-UNICAMP, CP 6165, 13083-970, Campinas, SP, Brazil}
\newcommand{\UFG}{Instituto de F\'{i}sica, Universidade Federal de Goi\'{a}s, CP 131, 74001-970, Goi\^{a}nia, GO, Brazil}
\newcommand{\USP}{Instituto de F\'{i}sica, Universidade de S\~{a}o Paulo,  CP 66318, 05315-970, S\~{a}o Paulo, SP, Brazil}
\newcommand{\Warsaw}{Department of Physics, University of Warsaw, Ho\.{z}a 69, PL-00-681 Warsaw, Poland}
\newcommand{\Washington}{Physics Department, Western Washington University, Bellingham, Washington 98225, USA}
\newcommand{\WandM}{Department of Physics, College of William \& Mary, Williamsburg, Virginia 23187, USA}
\newcommand{\Wisconsin}{Physics Department, University of Wisconsin, Madison, Wisconsin 53706, USA}
\newcommand{\deceased}{Deceased.}

\affiliation{\ANL}
\affiliation{\Athens}
%\affiliation{\Benedictine}
\affiliation{\BNL}
\affiliation{\Caltech}
\affiliation{\Cambridge}
\affiliation{\UNICAMP}
%\affiliation{\CdF}
\affiliation{\FNAL}
\affiliation{\UFG}
\affiliation{\Harvard}
\affiliation{\HolyCross}
\affiliation{\IIT}
\affiliation{\Indiana}
\affiliation{\Iowa}
%\affiliation{\IHEP}
%\affiliation{\ITEP}
%\affiliation{\JMU}
%\affiliation{\Lebedev}
%\affiliation{\LLL}
\affiliation{\UCL}
\affiliation{\Minnesota}
\affiliation{\Duluth}
\affiliation{\Otterbein}
\affiliation{\Oxford}
\affiliation{\Pittsburgh}
\affiliation{\RAL}
\affiliation{\USP}
\affiliation{\Carolina}
\affiliation{\Stanford}
\affiliation{\Sussex}
\affiliation{\TexasAM}
\affiliation{\Texas}
\affiliation{\Tufts}
\affiliation{\Warsaw}
%\affiliation{\Washington}
\affiliation{\WandM}
%\affiliation{\Wisconsin}

\author{P.~Adamson}
\affiliation{\FNAL}
%\affiliation{\UCL}
%\affiliation{\Sussex}

%\author{C.~Andreopoulos}
%\affiliation{\RAL}
%\affiliation{\Athens}

%\author{K.~E.~Arms}
%\affiliation{\Minnesota}

%\author{R.~Armstrong}
%\affiliation{\Indiana}

%\author{D.~J.~Auty}
%\affiliation{\Sussex}

%\author{S.~Avvakumov}
%\affiliation{\Stanford}

\author{D.~S.~Ayres}
\affiliation{\ANL}

\author{C.~Backhouse}
\affiliation{\Oxford}

%\author{B.~Baller}
%\affiliation{\FNAL}

%\author{B.~Barish}
%\affiliation{\Caltech}

%\author{P.~D.~Barnes~Jr.}
%\affiliation{\LLL}

\author{G.~Barr}
\affiliation{\Oxford}

%\author{W.~L.~Barrett}
%\affiliation{\Washington}

%\author{E.~Beall}
%\altaffiliation[Now at\ ]{\Cleveland .}
%\affiliation{\ANL}
%\affiliation{\Minnesota}

%\author{B.~R.~Becker}
%\affiliation{\Minnesota}

%\author{A.~Belias}
%\affiliation{\RAL}

%\author{R.~H.~Bernstein}
%\affiliation{\FNAL}

%\author{M.~Betancourt}
%\affiliation{\Minnesota}

%\author{D.~Bhattacharya}
%\affiliation{\Pittsburgh}

%\author{M.~Bhattarai}
%\affiliation{\Texas}
%\affiliation{\Duluth}

\author{M.~Bishai}
\affiliation{\BNL}

\author{A.~Blake}
\affiliation{\Cambridge}

%\author{B.~Bock}
%\affiliation{\Duluth}

\author{G.~J.~Bock}
\affiliation{\FNAL}

\author{D.~J.~Boehnlein}
\affiliation{\FNAL}

\author{D.~Bogert}
\affiliation{\FNAL}

%\author{P.~M.~Border}
%\affiliation{\Minnesota}

%\author{C.~Bower}
%\affiliation{\Indiana}

%\author{E.~Buckley-Geer}
%\affiliation{\FNAL}

\author{S.~V.~Cao}
\affiliation{\Texas}

%\author{S.~Cavanaugh}
%\affiliation{\Harvard}

%\author{J.~D.~Chapman}
%\affiliation{\Cambridge}

%\author{D.~Cherdack}
%\affiliation{\Tufts}

% ...

\collaboration{The MINOS Collaboration}
\noaffiliation

%% file: paper.bbl
\begin{thebibliography}{30}
\expandafter\ifx\csname natexlab\endcsname\relax\def\natexlab#1{#1}\fi
\expandafter\ifx\csname bibnamefont\endcsname\relax
  \def\bibnamefont#1{#1}\fi
\expandafter\ifx\csname bibfnamefont\endcsname\relax
  \def\bibfnamefont#1{#1}\fi
\expandafter\ifx\csname citenamefont\endcsname\relax
  \def\citenamefont#1{#1}\fi
\expandafter\ifx\csname url\endcsname\relax
  \def\url#1{\texttt{#1}}\fi
\expandafter\ifx\csname urlprefix\endcsname\relax\def\urlprefix{URL }\fi
\providecommand{\bibinfo}[2]{#2}
\providecommand{\eprint}[2][]{\url{#2}}

\bibitem[{ref({\natexlab{a}})}]{ref:minosRHC2011}
\bibinfo{note}{P.~Adamson et al. (MINOS), Phys. Rev. Lett. {\bf 107}, 021801
  (2011)}.

\bibitem[{ref({\natexlab{b}})}]{ref:minos2006}
\bibinfo{note}{D.~G.~Michael et al. (MINOS), Phys. Rev. Lett. {\bf 97}, 191801
  (2006)}.

\bibitem[{ref({\natexlab{c}})}]{ref:minos2008}
\bibinfo{note}{P.~Adamson et al. (MINOS), Phys. Rev. D {\bf 77}, 072002
  (2008)}.

\bibitem[{ref({\natexlab{d}})}]{ref:minosCC2010}
\bibinfo{note}{P.~Adamson et al. (MINOS), Phys. Rev. Lett. {\bf 106}, 181801
  (2011)}.

\bibitem[{\citenamefont{Ashie et~al.}(2004)}]{ref:sk}
\bibinfo{author}{\bibfnamefont{Y.}~\bibnamefont{Ashie}} \bibnamefont{et~al.}
  (\bibinfo{collaboration}{Super-Kamiokande}), \bibinfo{journal}{Phys. Rev.
  Lett.} \textbf{\bibinfo{volume}{93}}, \bibinfo{pages}{101801}
  (\bibinfo{year}{2004}).

\bibitem[{ref({\natexlab{e}})}]{ref:soudan2}
\bibinfo{note}{W.~W.~M.~Allison et al. (Soudan-2), Phys. Rev. D {\bf 72},
  052005 (2005)}.

\bibitem[{ref({\natexlab{f}})}]{ref:macro}
\bibinfo{note}{M.~Ambrosio et al. (MACRO), Eur. Phys. J. C. {\bf 36}, 323
  (2004)}.

\bibitem[{\citenamefont{Ahn et~al.}(2006)}]{ref:k2k}
\bibinfo{author}{\bibfnamefont{M.~H.} \bibnamefont{Ahn}} \bibnamefont{et~al.}
  (\bibinfo{collaboration}{K2K}), \bibinfo{journal}{Phys. Rev. D}
  \textbf{\bibinfo{volume}{74}}, \bibinfo{pages}{072003}
  (\bibinfo{year}{2006}).

\bibitem[{\citenamefont{Ahmad et~al.}(2002)}]{ref:sno}
\bibinfo{author}{\bibfnamefont{Q.~R.} \bibnamefont{Ahmad}} \bibnamefont{et~al.}
  (\bibinfo{collaboration}{SNO}), \bibinfo{journal}{Phys. Rev. Lett.}
  \textbf{\bibinfo{volume}{89}}, \bibinfo{pages}{011301}
  (\bibinfo{year}{2002}).

\bibitem[{\citenamefont{Abe et~al.}(2008)}]{ref:kamland}
\bibinfo{author}{\bibfnamefont{S.}~\bibnamefont{Abe}} \bibnamefont{et~al.}
  (\bibinfo{collaboration}{KamLAND}), \bibinfo{journal}{Phys. Rev. Lett.}
  \textbf{\bibinfo{volume}{100}}, \bibinfo{pages}{221803}
  (\bibinfo{year}{2008}).

\bibitem[{ref({\natexlab{g}})}]{ref:PMNS}
\bibinfo{note}{Z.~Maki, M.~Nakagawa, and S.~Sakata, Prog Theor. Phys. {\bf 28},
  870 (1962); B.~Pontecorvo, Zh. Eskp. Teor. Fiz. {\bf 53}, 1717 (1967) [Sov.
  Phys. JETP {\bf 26}, 984 (1968)]; V.~N.~Gribov and B.~Pontecorvo, Phys. Lett.
  B {\bf 28}, 493 (1968)}.

\bibitem[{\citenamefont{Mohapatra and Senjanovic}(1980)}]{Mohapatra:1979ia}
\bibinfo{author}{\bibfnamefont{R.~N.} \bibnamefont{Mohapatra}}
  \bibnamefont{and}
  \bibinfo{author}{\bibfnamefont{G.}~\bibnamefont{Senjanovic}},
  \bibinfo{journal}{Phys. Rev. Lett.} \textbf{\bibinfo{volume}{44}},
  \bibinfo{pages}{912} (\bibinfo{year}{1980}).

\bibitem[{\citenamefont{Fukugita and Yanagida}(1986)}]{Fukugita:1986hr}
\bibinfo{author}{\bibfnamefont{M.}~\bibnamefont{Fukugita}} \bibnamefont{and}
  \bibinfo{author}{\bibfnamefont{T.}~\bibnamefont{Yanagida}},
  \bibinfo{journal}{Phys. Lett. B} \textbf{\bibinfo{volume}{174}},
  \bibinfo{pages}{45} (\bibinfo{year}{1986}).

\bibitem[{ref({\natexlab{h}})}]{ref:PDG}
\bibinfo{note}{K. Nakamura et al. (Particle Data Group), J. Phys. G {\bf 37},
  075021 (2010)}.

\bibitem[{\citenamefont{Parke}(2008)}]{Parke:2008zz}
\bibinfo{author}{\bibfnamefont{S.~J.} \bibnamefont{Parke}}, in
  \emph{\bibinfo{booktitle}{{Neutrino oscillations: Present status and future
  plans}}}, edited by \bibinfo{editor}{\bibfnamefont{J.~A.}
  \bibnamefont{Thomas}} \bibnamefont{and} \bibinfo{editor}{\bibfnamefont{P.~L.}
  \bibnamefont{Vahle}} (\bibinfo{publisher}{World Scientific},
  \bibinfo{year}{2008}), \bibinfo{note}{also published as
  FERMILAB-PUB-07-767-T}.

\bibitem[{\citenamefont{Araki et~al.}(2005)}]{ref:kamland2}
\bibinfo{author}{\bibfnamefont{T.}~\bibnamefont{Araki}} \bibnamefont{et~al.}
  (\bibinfo{collaboration}{KamLAND}), \bibinfo{journal}{Phys. Rev. Lett.}
  \textbf{\bibinfo{volume}{94}}, \bibinfo{pages}{081801}
  (\bibinfo{year}{2005}).

\bibitem[{ref({\natexlab{i}})}]{ref:mattereffect}
\bibinfo{note}{In matter $P(\numu\rightarrow\numu)$ and
  $P(\numubar\rightarrow\numubar)$ can differ by as much as 0.1\% due to
  $\numu{} \leftrightarrow \nue{}$ mixing and $\nue{}$ and $\nuebar{}$
  scattering on electrons. This effect is too small to be observed in our
  experiment.}

\bibitem[{\citenamefont{Wolfenstein}(1978)}]{Wolfenstein:1977ue}
\bibinfo{author}{\bibfnamefont{L.}~\bibnamefont{Wolfenstein}},
  \bibinfo{journal}{Phys. Rev. D} \textbf{\bibinfo{volume}{17}},
  \bibinfo{pages}{2369} (\bibinfo{year}{1978}).

\bibitem[{\citenamefont{Michael et~al.}(2008)}]{ref:minosnim}
\bibinfo{author}{\bibfnamefont{D.~G.} \bibnamefont{Michael}}
  \bibnamefont{et~al.} (\bibinfo{collaboration}{MINOS}),
  \bibinfo{journal}{Nucl. Instrum. Meth. A} \textbf{\bibinfo{volume}{596}},
  \bibinfo{pages}{190} (\bibinfo{year}{2008}).

\bibitem[{ref({\natexlab{j}})}]{ref:beamcomp}
\bibinfo{note}{Accounting for flux and cross-sections, the \numu{}(\numubar{})
  enhanced beam consists of 91.7\%(58.1\%) \numu{}, 7.0\%(39.9\%) \numubar{}
  and 1.3\%(2.0\%) \nue{}+\nuebar{}}.

\bibitem[{\citenamefont{Abe et~al.}(2012)}]{ref:t2k}
\bibinfo{author}{\bibfnamefont{K.}~\bibnamefont{Abe}} \bibnamefont{et~al.}
  (\bibinfo{collaboration}{T2K}) (\bibinfo{year}{2012}),
  \eprint{arXiv:hep-ex/1201.1386}.

\bibitem[{\citenamefont{Ospanov}(2008)}]{ref:RustemThesis}
\bibinfo{author}{\bibfnamefont{R.}~\bibnamefont{Ospanov}},
  \bibinfo{journal}{Ph.D. Thesis, University of Texas at Austin}
  (\bibinfo{year}{2008}).

\bibitem[{\citenamefont{Backhouse}(2011)}]{ref:ChrisThesis}
\bibinfo{author}{\bibfnamefont{C.}~\bibnamefont{Backhouse}},
  \bibinfo{journal}{D.Phil. Thesis, Oxford University}  (\bibinfo{year}{2011}).

\bibitem[{ref({\natexlab{k}})}]{ref:beam}
\bibinfo{note}{S.~Kopp, arXiv:physics/0508001}.

\bibitem[{\citenamefont{Anderson et~al.}(1998)}]{ref:beam2}
\bibinfo{author}{\bibfnamefont{K.}~\bibnamefont{Anderson}}
  \bibnamefont{et~al.}, \bibinfo{type}{Tech. Rep.},
  \bibinfo{institution}{Fermilab} (\bibinfo{year}{1998}),
  \bibinfo{note}{{FERMILAB-DESIGN-1998-01}}.

\bibitem[{ref({\natexlab{l}})}]{ref:horns}
\bibinfo{note}{A zero field is expected due to symmetry around the beam axis.
  We measure a $\unit[1\times 10^{-2}]{mrad}$ deflection of the primary proton
  beam, corresponding to $B\cdot \mathrm{d}\ell=\unit[43]{G m}$.}

\bibitem[{ref({\natexlab{m}})}]{ref:spilltrigger}
\bibinfo{note}{The window is defined by GPS time-stamping each beam extraction
  and transmitting the time stamp over the internet to the FD, which buffers
  data to cope with latency.}

\bibitem[{\citenamefont{Dytman et~al.}(2008)\citenamefont{Dytman, Gallagher,
  and Kordosky}}]{ref:neugen}
\bibinfo{author}{\bibfnamefont{S.}~\bibnamefont{Dytman}},
  \bibinfo{author}{\bibfnamefont{H.}~\bibnamefont{Gallagher}},
  \bibnamefont{and} \bibinfo{author}{\bibfnamefont{M.}~\bibnamefont{Kordosky}}
  (\bibinfo{year}{2008}), \eprint{arXiv:hep-ex/0806.2119}.

\bibitem[{ref({\natexlab{n}})}]{ref:MesonExchangeCurrents}
\bibinfo{note}{A.~Bodek, H. S.~Budd and M. E.~Christy, Eur.\ Phys.\ J. C {\bf
  71}, 1726 (2011)}.

\bibitem[{ref({\natexlab{o}})}]{ref:FC}
\bibinfo{note}{G. J. Feldman and R. D. Cousins, Phys. Rev. D {\bf 57}, 3873
  (1998)}.

\end{thebibliography}
